\documentstyle[12pt,epsf]{article}
\def\be{\begin{equation}}
\def\ee{\end{equation}}
\renewcommand{\theequation}{\thesection.\arabic{equation}}
\begin{document}

\begin{titlepage}

\begin{flushright}

BRX TH-428 \\
ULB-TH-97/23\\
IASSNS-HEP-97138

\end{flushright}

\begin{center}
{\large\bf  $p$-Brane Dyons and
Electric-magnetic Duality}

\end{center}
\vfill

\begin{center}
{\large
S. Deser$^{a}$,
A. Gomberoff$^{b}$,
M. Henneaux$^{b,c }$ \\ and
C. Teitelboim$^{b,d \ *}$}
\end{center}
\vfill

\begin{center}{\sl
$^a$ Department of Physics, Brandeis University,\\
Waltham, MA 02254, U.S.A.\\[1.5ex]

$^b$ Centro de Estudios Cient\'\i ficos de Santiago,\\
Casilla 16443, Santiago 9, Chile\\[1.5ex]

$^c$ Facult\'e des Sciences, Universit\'e Libre de
Bruxelles,\\
Campus Plaine C.P. 231, B--1050 Bruxelles, Belgium\\[1.5ex]

$^d$ Institute for Advanced Study,\\
Princeton, New Jersey 08540, U.S.A.

}\end{center}
\vfill
%\break

\begin{abstract}
We discuss dyons, charge quantization and electric-magnetic duality
for self-interacting, abelian, $p$--form theories in the spacetime dimensions
$D=2(p+1)$ where dyons can be present. The corresponding quantization
conditions and duality properties
are strikingly different depending on whether $p$ is odd or even. If $p$ is
odd
one has the familiar $e\bar{g}-g\bar{e}=2\pi n \hbar$, 
whereas for even $p$ one finds the opposite relative sign,
$e\bar{g}+g\bar{e}=2\pi n \hbar$. These conditions are obtained
by introducing Dirac strings and taking due account of the multiple
connectedness of the configuration space of the strings and the dyons. A 
two-potential formulation of the theory that treats the electric and magnetic
sources on the same footing is also given. Our results hold for arbitrary
gauge invariant self-interaction of the fields and are valid 
irrespective of their duality properties. 
\end{abstract}
\vfill
\vskip 0.2cm
\rule{3cm}{0.1mm}

\noindent 
\footnotesize{{\tt deser@binah.cc.brandeis.edu,  andy@cecs.cl,  henneaux@ulb.ac.be, teitel@cecs.cl}}
\end{titlepage}

%%%%%%%%%%%%%%%%%%%%%%%%%%%%%%%%%%%%%%%%%%%%%%%%%%%%%
\section{Introduction}
%%%%%%%%%%%%%%%%%%%%%%%%%%%%%%%%%%%%%%%%%%%%%%%%%%%%%

\setcounter{equation}{0}

Ever since Dirac\cite{Dirac} introduced it in quantum mechanics in 1931, the
magnetic pole has been a fascinating object.  In particular, the
quantization condition relating the electric and magnetic
charges, its relationship with the symmetry between electricity and magnetism,
and the possible generalizations of Dirac's approach, remain questions of high
interest today.  Among the natural extensions of Dirac's own improved
formulation\cite{Dirac2} 
of 1948, two will be of importance here. The first is the consideration of
dyons, particles having both electric and magnetic
charge\cite{Schwinger,Zwan}.
The second is the generalization to higher dimensions where the
electromagnetic
potential is replaced by a $p$--form and the electric charges become extended
objects with a $p$--dimensional history instead of a worldline\cite{CT,Nepo}.
We will establish the following results, some of which were briefly discussed 
or  implicit in \cite{DGHT}:

\begin{enumerate}

\item The quantization condition for the electric and magnetic charges is
present even when the theory is not duality invariant (this was to be expected
since -- as shown in \cite{CT,Nepo} -- the quantization condition holds even
when the electric and magnetic charges are extended objects of different --
complementary -- dimensions).

\item For spacetime dimensions $D=2(p+1)$ such that dyons can exist, 
the quantization condition takes different forms depending on the parity of
$p$,
namely
\begin{eqnarray}
e\bar{g}-g\bar{e}&=&2\pi n \hbar \ ,
D=4k \ (p\mbox{ odd}) \ \ , \label{minus}\\
e\bar{g}+g\bar{e}&=&2\pi n \hbar \ ,
D=4k+2 \ (p\mbox{ even})\ \ \label{plus} .
\end{eqnarray}
These conditions are obtained by  using Dirac strings and analyzing the
connectivity of the configuration space of the dyons and the strings. They are
also shown to arise from the generalization to higher dimensions of the 
arguments based on:  (i) the quantization of the angular momentum stored
in the field of electric and magnetic poles\cite{Fierz} and (ii)
compatibility of regular 
local gauge charts\cite{WY}.

\item A two--potential formulation 
(which is not manifestly Lorentz invariant)
can be given for any spacetime dimension. 
For, and only for, $D=4k$ there are special self--interactions for which the 
source-free theory
is invariant under duality rotations; there the 
two--potential formulation exhibits duality invariance in a particular
transparent way as a normal Noether
symmetry.

\end{enumerate}

Although most of the results discussed here were first obtained using the 
two-potential formulation which permits a more economical analysis, we will 
also present them in terms of the more familiar one-potential representation.

The plan of the paper is as follows. Section 2 presents an extension of
Dirac's
1948 formulation\cite{Dirac2} for electrodynamics with magnetic poles which 
allows for dyons and also for self--interactions of the electromagnetic
field. This is the prototype of the $p$--form theory for $p$ odd and $D=2(p+1)=4k$.
The dual formulation where a potential is introduced for the electric field
rather than for the magnetic field is also discussed. Section 3 discusses
the prototype theory with even $p$ and $D=2(p+1)=4k+2$. This is the case
$p=2$, $D=6$.
Section 4 then treats the two-potential formulation of both the odd and even 
$p$ cases, ending with a brief discussion of the
generalization to higher dimensions.  Some other aspects of interest are 
included in the appendices:  Appendix A discusses the issues of topology 
of configuration space and orientation of surfaces 
which have bearing on the quantization condition. Appendix B discusses 
the general solution of the quantization conditions in
the symmetric case $e\bar{g}+g\bar{e}=2\pi n
\hbar$ , Appendix C gives the analog of the angular momentum  
quantization conditions, and Appendix D discusses the argument based on
compatibility of local gauge charts.

%%%%%%%%%%%%%%%%%%%%%%%%%%%%%%%%%%%%%%%%%%%%%%%%%%%%%%%%%%%%%%%%%%
\section{Electrodynamics in D=4; Dyons}
%%%%%%%%%%%%%%%%%%%%%%%%%%%%%%%%%%%%%%%%%%%%%%%%%%%%%%%%%%%%%%%%%%%

\subsection{Action}
%%%%%%%%%%%%%%%%%%%%%%

The covariant action describing the coupling of dyons
to non-linear electrodynamics in four dimensions
is \cite{Dirac2}
\begin{equation}
I[A_\mu, z_n, y_n] = I_F + I_C+ I_{P}
\label{action}
\end{equation}
with
\begin{eqnarray}
I_F &=& \int d^4x \, {\cal L}(F_{\mu \nu}), \\
I_C &=& \sum_n e_n \int A_\mu (z_n) dz^\mu_n,
\label{C}\\
I_{P} &=& -  \sum_n \: m_n \int \sqrt{-(dz^\mu )^2} .
\end{eqnarray}

The field strength $F_{\mu \nu}$ is
defined through
\begin{equation}
F_{\mu \nu} = \partial_\mu A_\nu - \partial_\nu A_\mu
+  ^{*} \! G_{\mu \nu}
\label{F}
\end{equation}
with
\begin{eqnarray}
^*G_{\mu \nu} &=& \frac{1}{2} \epsilon_{\mu \nu \alpha \beta}
G^{\alpha \beta}, \\
G^{\mu \nu} &=& \sum_n g_n \int dy^\mu_n \wedge dy^\nu_n
\delta^4 (x-y_n).
\label{stringcontribution}
\end{eqnarray}

Here we call $(e_n,g_n,m_n)$ the electric charge, magnetic charge and mass of
the $n$--th particle respectively. We use signature (-,+,+,+)
and take $\epsilon_{0123}=1$.  The electric
and magnetic couplings are described asymmetrically:
only the magnetic charge appears in the definition (\ref{F}) of
the electromagnetic strength $F_{\mu \nu}$, and only the electric
charge enters the minimal coupling term (\ref{C}).

We attach to each particle a string $y_n(\sigma_n,\tau_n )$, with
$0 \le \sigma_n <\infty$ and $-\infty < \tau_n < \infty$.
A particle trajectory is specified by
\mbox{$z^\mu_n (\tau_n ) = y^\mu (\sigma_n = 0, \;\tau_n )$}.

The equations of motion
that follow from
(\ref{action}) are
\begin{eqnarray}
\partial_\mu H^{\mu \nu} &=& - j^\nu_e, \label{Maxwell} \\
m_n \ddot{ z}_n ^\mu &=& (e_n F^\mu_{\,\nu} + g_n \,
 ^{*} \! H^\mu_{\,\nu})
\dot{ z}_n ^\nu
\label{Lorentz}
\end{eqnarray}
where $j^\nu_e$ is the electric current,
\be
j^\nu_e = \sum_n e_n \int dz^\mu_n \delta^4(x-z_n),
\ee
and where the evolution parameter $\tau_n$ in (\ref{Lorentz})
is the particle's proper time.  We have defined
\be
H^{\mu \nu} = - 2 \frac{\partial {\cal L}}{\partial F_{\mu\nu}}.
\ee
For Maxwell theory ${\cal L} = (-1/4) F^{\mu \nu}F_{\mu \nu}$ and
$H^{\mu \nu} = F^{\mu \nu}$. Equations (\ref{Maxwell}) and (\ref{Lorentz})
arise from extremization of the action with respect to the vector potential
$A_\mu$ and the particles coordinates $z^\mu_n$, respectively. Extremization 
of the action with respect to the string coordinates yields no
equation provided the string attached to particle $n$ passes through
no other particle (``Dirac veto"). This remains true even if each particle has
both electric and magnetic charge. The original Dirac veto stated that the 
Dirac string of a magnetic pole cannot pass through an electric charge. 
One can verify that the argument remains valid for dyons in the form just 
stated.  The analysis follows, step by step, that of \cite{Dirac2}. 
The only difference is the appearance of a term  of the form 
\begin{eqnarray}
&& \epsilon_{\mu \nu \alpha \beta} 
\int d^4x \int
dy^\mu
\wedge dy^\nu \int dz^\alpha  \delta y^\beta
\delta^4(x-y) \delta^4(x-z) = \label{singular}\\
&& \epsilon_{\mu \nu \alpha \beta} \int d^4x \int d\tau d\sigma
d\tau' (\dot{y}^\mu y'^\nu
- \dot{y}^\nu y'^\mu ) \dot{z}^\alpha \delta y^\beta
\delta^4(x-y(\tau,
\sigma)) \delta^4(x-z(\tau')) \nonumber 
\end{eqnarray}
in the variation of the action with respect to the string coordinates. However
(\ref{singular})  actually vanishes:  the integral over $x^\mu$
forces $y^\mu = z^\mu$, {\it i.e.}, $\sigma = 0$ and $\tau =\tau'$,
implying $\dot{y}^\mu_n = \dot{z}^\mu_n$.
Thus, the whole expression is actually zero because
$\epsilon_{\mu \nu \alpha \beta}$ is antisymmetric in
$\mu$, $\alpha$, while the product $\dot{z}^\mu \dot{z}^\alpha$
is symmetric.  Whenever necessary, we shall therefore regularize such
expressions to zero [The integrand is formally singular because it involves
$\delta^2(0)$]. The magnetic current appears as a ``source for the 
Bianchi identity'',
\be
\partial_\mu \,  ^{*} \! F^{\mu \nu} = j^\nu_m
\ee
with
\be
j^\nu_m = \sum_n g_n \int dz^\mu_n \delta^4(x-z_n),
\ee

\subsection{Charge Quantization Condition}
%%%%%%%%%%%%%%%%%%%%%%%%%%%%%%%%%%%%%%

We now derive the quantization condition for the
dyon charges, following Dirac's argument\cite{Dirac2} for ``pure" sources
based on the unobservability of the strings.  The extension
to dyons presents no difficulty if one takes proper
account of the multiple--connectedness of the configuration
space of the strings and charges: 
We shall give it in detail here, 
not having seen it in the literature.

It is convenient to choose as time coordinate on the strings
the zeroth coordinate $y^0$ itself, $\tau = y^0$.
The momenta conjugate to the
string spatial coordinates are then constrained by
\be
\pi_{i} = -g  ^* \! H_{ik} \frac{\partial y^k}{\partial \sigma} .
\label{mom}
\ee

The dependence of the wave functional $\Psi[A_i, z_n, y_n]$ on the
string coordinates is entirely determined by the constraints.
This reflects, in the quantum theory, the fact that the strings
carry no degree of freedom of their own. The dependence of $\Psi$
on $y_n$ follows from integrating the quantum constraints,
\be
\frac{\hbar}{i} \frac{\delta \Psi}{\delta y_n^i}
=  -\left[g_n  ^* \! H_{ik} \frac{\partial y_n^k}{\partial \sigma}\right] \Psi
\label{magnetic}
\ee
in the configuration space of the string coordinates, the particle
coordinates and the vector potential. This space is not simply connected, 
because of the Dirac veto, and the general requirement is: Circling a
loop in configuration space which is contractible to a point, the
wave function must return to itself. If the loop is not contractible the wave
functions need not be single--valued, but must just form a representation of
the fundamental homotopy group $\pi_1$\cite{???}.

\begin{figure}
\begin{center}
  \leavevmode
 \epsfbox{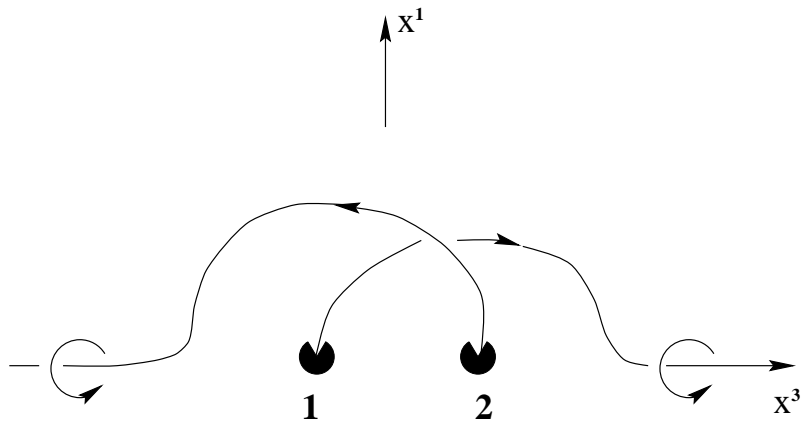}
\label{f1}
\caption{Double pass.  The string of dyon 1 goes towards
dyon 2 without touching it, whereas that of dyon 2 goes towards dyon 1 again
without touching it. The strings themselves  do not touch either.
Rather,
string 1 is behind (more into the page than string 2). We now imagine a
simultaneous full turn of both strings going out of the page in the
$(x^1,x^2)$
plane orthogonal to the $x^3$  axis. The dyons are kept fixed. This ``motion''
describes a path in the configuration space of the dyons and the strings which
we call ``the double pass''. The fact that string 1 does not touch dyon 2,
and 
vice versa, is mandatory (Dirac
veto). However, in the double pass we also require the strings  not to touch
each other at any ``instant'' during the motion that generates it. This is not
mandatory in general, but it is what makes the double pass contractible to a
point.}
\end{center}
\end{figure}

Consider the double-pass motion shown in figure 1, and discussed in detail in 
Appendix A. The string attached to dyon 1, with charges $(e,g)$, rotates 
around dyon 2, with charges $(\bar{e},\bar{g})$, while the
string attached to 2 rotates at the same time around 1.
The two strings rotate out of the sheet and the first string is
behind the second string, so that they never touch.
The simultaneous $2 \pi$ turn of both strings
(``double-pass") is a closed
contractible loop in the configuration space of the
strings {\it and} the particles. By
contrast, the turn of a single string
is {\it not} contractible in the space defined by the vetos.

Since the double-pass is contractible, the phase picked up
by the wave function should be a multiple of $2 \pi$.
In that motion, the rotation of the string attached
to the first dyon brings in the  phase
$(1/ \hbar) g \bar{e}$ because of (\ref{magnetic}) and the Gauss constraint
$\partial_j E^j = j^0_e$.
Similarly, the
rotation of the string attached to the second
dyon brings in  the phase
$- (1/ \hbar)e \bar{g}$, with a minus sign
because the orientations of the two-surfaces
swept out by the strings in their turning are opposite.
Hence, the total phase
is equal to $ (1/ \hbar)(\bar{e} g - e \bar{g})$.
Single-valuedness of the wave function for this
contractible motion implies the Dirac quantization condition 
\cite{Dirac},

\be
\bar{e} g - e \bar{g} = 2 \pi n \, \hbar \ .
\label{eg}
\ee

Note that if the first dyon is purely magnetic
($e = 0$) and the second purely electric ($\bar{g}=0$)
-- the case considered by Dirac --,
the phase is entirely accounted for by the motion of the
string attached to the magnetic pole.  In fact, the
constraint (\ref{magnetic}), which involves only the magnetic couplings,
obviously implies that the wave function
should not depend on the coordinates of strings attached to
purely electric particles since it reduces to
$\delta \Psi /\delta y^i_n = 0$ (for particle $n$ purely
electric).  It is thus
clear that no non-trivial phase can be generated by a motion of these
strings.  One can thus drop them altogether, in agreement
with the original treatment by Dirac in which no string was ever
attached to purely
electrically charged particles.   In the configuration space where the electric poles carry
no string, the motion of figure 1 -- in which there is now
only one string -- becomes contractible,
because one can now move the magnetic pole without restriction [the
restriction that it must avoid the string attached to the electric
particle is absent].
Thus it is legitimate -- and actually mandatory -- to require
that the phase $(1/\hbar) g \bar{e}$ associated with that single motion
should be a multiple of
$2 \pi$, in agreement with what (\ref{eg}) reduces to in
this case.

By contrast, it would be wrong
to require that the phase picked up by the wave function in the
rotation of a single string is a multiple of $2 \pi$ for generic dyons
because that motion is not homotopic to the trivial motion:
only the double-pass is.  The phase around a non-contractible
loop depends in fact on the chosen representation and is equal, as we have
seen, to $(1/\hbar) g \bar{e}$ or $-(1/\hbar) e \bar{g}$, which
are in general not integer multiples of $2 \pi$ (only (\ref{eg})
holds).

\subsection{Dual formulation}
%%%%%%%%%%%%%%%%%%%%%%%%%%%%%%%%%%%

\label{dual4}

In the above treatment, there is a vector potential for $F_{\mu \nu}$
but none for $^{*} \! H_{\mu \nu}$.  One may go to an
alternative representation
in which the roles of $F_{\mu \nu}$ and $^{*} \! H_{\mu \nu}$ are
exchanged, as well as the roles of magnetic and electric
charges.  This may be done for all values of the rank of the $p$-form and the
spacetime
dimension $D$. We describe here the procedure for $p=1$, $D=4$. A Lagrange
multiplier $S_{\alpha \beta}$ is introduced
for the Bianchi identity, i.e. for the definition (\ref{F}) of $F_{\mu
\nu}$, so that the action becomes
\be
I[A, F, S, z, y] = I_F + I_{C}+ I_{P}
- \frac{1}{4} \int d^4x \epsilon^{\alpha \beta \mu \nu}
S_{\alpha \beta} ( F_{\mu \nu} - \partial_\mu A_\nu +
\partial_\nu A_\mu - ^{*} \! G_{\mu \nu}).
\label{LagMult}
\ee
We have adjusted the coefficient of the Lagrange
multiplier term so that $S_{\alpha \beta} =
^{*} \! H_{\alpha \beta}$ follows from the $F$-equation of
motion.
If one expresses $F$ in terms of $S$ from this equation
(we assume this to be possible), one gets the action
\be
I[A, S, z, y] = I_S + \frac{1}{2} \int d^4x [
\epsilon^{\alpha \beta \mu \nu}
S_{\alpha \beta} \partial_\mu A_\nu -  S^{\mu \nu}
G_{\mu \nu}] + I_{C}+ I_{P}
\label{actionn}
\ee
where
\be
I_S = \int d^4x [{\cal L}(F(S)) - \frac{1}{4} \epsilon^{\alpha \beta \mu \nu}
S_{\alpha \beta}  F_{\mu \nu}(S)] \equiv \int d^4x \bar{{\cal L}}(S).
\label{actionS}
\ee
The original vector potential $A_\mu$ appears linearly in (\ref{actionn})
and may be viewed as a Lagrange multiplier for the Bianchi identity for
$S_{\mu \nu}$,
\be
\frac{1}{2} \epsilon^{\alpha \beta \mu \nu} \partial_\beta S_{\mu \nu}
+ j^\alpha_e = 0.
\label{BianchiS}
\ee
If one solves (\ref{BianchiS}) by expressing $S_{\mu \nu}$ as the exterior
derivative of a vector potential $Z_\mu$ plus a string term,
\be
S_{\mu \nu} = \partial_\mu Z_\nu - \partial_\nu Z_\mu + ^{*}\! M_{\mu \nu},
\label{defS}
\ee
with
\be
M^{\mu \nu} = \sum_n e_n \int dy^\mu_n \wedge dy^\nu_n
\delta^4 (x-y_n),
\ee
one finds the action
\be
I[Z,z,y] = I_S + I_{C}^{magn.} + I_P - \frac{1}{2}
\int d^4x ^{*}\! M_{\mu \nu} G^{\mu \nu}
\label{actionZzy}
\ee
where $S_{\mu \nu}$ is defined through (\ref{defS}) and the
magnetic minimal coupling term is
\be
I_{C}^{magn.} = - \sum_n g_n \int dz_n^\mu Z_\mu(z_n) \; .
\label{min}
\ee

The last term in the action (\ref{actionZzy}) can be written as the
integral of a total time derivative,
\be
\int d^4x ^{*}\! M_{\mu \nu} G^{\mu \nu}
= \int dt \frac{\delta V}{\delta y^\alpha_n} \dot{y}^\alpha_n
+ \int dt \frac{\delta V}{\delta z^\alpha_n} \dot{z}^\alpha_n
\label{total}
\ee
where $V$ is a functional of the coordinates of the strings and the
particles whose precise form will not be needed here.
That $V$ exists may be verified by observing that the
variational derivatives, keeping the endpoints fixed, of $\int d^4x ^{*}\!
M_{\mu \nu} G^{\mu \nu}$ all identically vanish, implying the form
(\ref{total}). [To check that these functional derivatives vanish, one
must use the fact that expressions like (\ref{singular}) above
are zero.]. However, since the configuration space of the 
particles and the
strings is multiply connected because of the veto, the
functional $V$, which exists locally, may be multiple-valued.
That is, the integral  of its (functional) gradient along
a non-contractible loop may be non-zero. Below, we will actually exhibit
a loop for which this is the case. One may add to the 
action the integral of the time
derivative of a multiply-valued functional.  This
does not modify the classical equations of motion, and
is also admissible quantum-mechanically, provided
one changes the representation of the fundamental
group in which the wave functions transform accordingly (see below). 
If one drops 
the last term in the action (\ref{actionZzy})
one gets the magnetic description of the interaction,
\be
I[Z,z,y] = I_S + I_{C}^{magn.} + I_P
\label{magneticDescri}
\ee
in which the electric minimal coupling to $A_\mu$ has
been replaced by a magnetic
minimal coupling to $Z_\mu$, and magnetic string terms
in $F$ have been replaced by electric string terms in the
definition of the dual field $S$.

When the theory is duality invariant, which happens when the
Lagrangian fulfills the conditions analysed in  \cite{GibbR}, the action
 $I_{S}$ is the same functional of $S$ as $I_F$ is of $F$. 
A particular example is given by the
Born-Infeld theory \cite{BornI}, whose Hamiltonian
has been studied from the duality point of view in
\cite{BB}-\cite{us} (the dual Lagrangian was explicitly
worked out by Born and Infeld, equation (4.12) of \cite{BornI}).

The purpose of the rather direct exercise of deriving
(\ref{magneticDescri}) from the original action is to
show that the phase picked up by the quantum wave functional
when integrating the constraints around non contractible
loops in configuration space
does depend on the (multiple-valued) functional
$V$.  Only the phase picked up around
contractible loops has an invariant meaning.

The quantum constraints read now
\be
\frac{\hbar}{i} \frac{\delta \Psi}{\delta y^i_{n}}
=\left[e_n F_{ik} \frac{\partial y_n^k}{\partial \sigma}\right] \Psi =-e_n
\epsilon_{ijk} B^j y'^k_n  \; \Psi
\label{electric}
\ee
where $B^i $
is the magnetic field,
\begin{eqnarray}
B^i &=& \frac{1}{2} \epsilon^{ijk} F_{jk} \nonumber \\
&=&  \epsilon^{ijk} \partial_j A_k + \sum_n g_n \int dy^0_n \wedge
dy^i_n \delta^{(4)}(x-y_n).
\label{B'}
\end{eqnarray}
Here, we have used
\be
\frac{\partial \bar{{\cal L}}}{S_{\mu \nu}} = -
\frac{1}{2} \, \,^{*}\! F^{\mu \nu}.
\ee
Consider the double-pass analyzed above.
When the first string, attached to the dyon $(e,g)$ turns
around the second dyon $(\bar{e}, \bar{g})$,
this time the integration of the constraint (\ref{electric}) brings in
the phase $-(1/\hbar)e \bar{g}$ rather than
$(1/\hbar)g  \bar{e}$.
One gets a phase different from that of above because the loop is not
contractible, indicating
that different representations of the fundamental group are involved
[Conversely, the existence of a closed one--form in configuration space
whose integral around the loop is not zero proves that the loop is not
contractible].  However, if one considers the combined motions involved in the
double-pass, which is contractible, one gets the same total phase
$(1/\hbar) (g  \bar{e} - e \bar{g})$ because the rotation
of the second string brings in now $(1/\hbar)g  \bar{e}$.
The quantization condition is, properly, unchanged.

%%%%%%%%%%%%%%%%%%%%%%%%%%%%%%%%%%%%%%%%%%%%%%%%%%%%%%%%%%%%%%%%%%%%%
\section{2-Forms in D=6;  Dyons}
%%%%%%%%%%%%%%%%%%%%%%%%%%%%%%%%%%%%%%%%%%%%%%%%%%%%%%%%%%%%%%%%%%%%%
\setcounter{equation}{0}

\subsection{Action}
%%%%%%%%%%%%%%%%%%%%%

Dyonic sources for $p$-forms can exist in spacetime
dimension $2(p+1)$. However, the quantization condition for 
the dyon charges of even $p$ is strikingly different
from that of odd $p$.

When one introduces $p$--branes which are the higher dimensional 
analogs of the Dirac
string one finds that the simplest contractible loop
for two dyons and their branes is the
higher-dimensional analog of the double-pass.  The new feature
for $p=2k$-forms is that the orientations of the surfaces
swept out by the two membranes are now the same, instead
of being opposite as in the $p=(2k+1)$--case.  Thus the total
phase for the double-pass acquires a relative plus sign instead of a
minus sign and one obtains $e\bar{g}+g\bar{e}=2\pi n \hbar$ as
the quantization condition instead of (\ref{eg}).
This change of orientation is explained in detail for $D=6$ dimensions in
Appendix A. There is an easier way to verify this, based on the symmetric
representation of the theory, which we describe in the next section.

The covariant action for the $2$-form coupled to dyons is,
\be
I[A_{\mu \nu}, z^\mu, \bar{z}^\mu, \bar{y}^\mu]
= I_F + I_{C} + I_P
\label{covariant}
\ee
with
\begin{eqnarray}
I_F &=&  \int d^6x {\cal L}(F^{\mu \nu \lambda})
, \label{if}\\
I_{C} &=& \sum_n \frac{e_n}{2} \int A_{\mu \nu}(z_n)
dz^\mu_n \wedge dz^\nu_n, \\
I_P &=& - \sum_n m_n \int (- ^{(2)}g_n)^{\frac{1}{2}} d^2\sigma_n.
\label{deter}
\end{eqnarray}
In (\ref{deter}), $g_n$ is the determinant of
the metric induced on the two-dimensional spacetime history
of the strings $z^\mu_n(\sigma^1_n, \sigma^2_n)$.
We attach a membrane with history $y_n^\mu(\sigma_n^1,
\sigma^2_n,\tau_n)$ to each string,
$y^\mu_n(\sigma^1_n,
\sigma^2_n, 0) = z^\mu_n(\sigma^1_n,
\sigma^2_n)$.  The field strength appearing in (\ref{if})
is,
\be
F_{\mu \nu \lambda}= \partial_\mu A_{\nu \lambda}
- \partial_\nu A_{\mu \lambda} - \partial_\lambda
A_{\nu \mu} + ^{*}\!G_{\mu \nu \lambda}
\ee
where $^{*}\!G_{\mu \nu \lambda}$ is the contribution of the
membranes,
\be
G^{\mu \nu \lambda} = \sum_n g_n \int dy^\mu_n \wedge
dy^\nu_n \wedge dy^\lambda_n \delta^6(x-y_n).
\ee
The equations of motion following from the action
(\ref{covariant}) consistently describe the coupled system
of the abelian $2$-form with the dyons,
provided the membrane attached to string $n$ does not
pass through the other strings (Dirac veto) \cite{CT}.
In particular, the $2$-form field equation and the Bianchi identity
are respectively
\begin{eqnarray}
\partial_\mu H^{\mu \nu \lambda} &=& - j^{\nu \lambda}_e, \\
\partial_\mu ^{*}\! F^{\mu \nu \lambda} &=&  j^{\nu \lambda}_m
\end{eqnarray}
with
\begin{eqnarray}
j^{\nu \lambda}_e &=& \sum_n e_n \int dz^\nu \wedge dz^\lambda
\delta^6(x-z_n)\\
j^{\nu \lambda}_m &=& \sum_n g_n \int dz^\nu \wedge dz^\lambda
\delta^6(x-z_n),
\end{eqnarray}
where
\be
H^{\mu \nu \lambda} = - (3)! \frac{\partial {\cal L}}{\partial
F_{\mu \nu \lambda}}
\ee
so that $H^{\mu \nu \lambda} = F^{\mu \nu \lambda}$ in the linear case.

\subsection{Charge Quantization}
%%%%%%%%%%%%%%%%%%%%%%%%%%%%%%%%%%%

One can now derive the quantization condition as in
$4$ dimensions.  Consider two dyons with
respective strengths $(e,g)$ and $(\bar{e}, \bar{g})$.
In the higher-dimensional generalization of the double-pass described
above, the membrane attached to the first dyon describes a three-dimensional
surface which links, in five-dimensional space, the other
dyonic string.  Simultaneously, the membrane attached to
the second dyon performs an analogous motion around the first
string.  Because, for each membrane,  the quantum constraints which are the analog of (\ref{magnetic}) read
\begin{equation}
\frac{\hbar}{i}\frac{\delta \Psi}{\delta y^i} = \left[g \ ^{*}\! H_{ijk} 
\frac{\partial y^j}{\partial \sigma^1}\frac{\partial y^k}{\partial \sigma^2}\right] \Psi \  ,
\label{qconstraint}
\end{equation}
one finds that the motion of the first membrane brings in the phase
$(1/\hbar) g\bar{e}$. 
The motion of the second string brings in
the phase $+(1/\hbar) \bar{g}e$ because the orientations are the same.
Therefore it is the plus sign which is realized, so
that the quantization condition is (\ref{plus})  as announced.

As for D=4 electrodynamics, there is a formulation dual to that of section
3.1. 
However, as explained in \cite{DGHT}, the hyperbolic duality
rotations that leave the equations of motion invariant do not respect
the action.
There is only a residual, discrete, duality that interchanges the
electric and magnetic fields, with plus signs. The theory is invariant 
under this discrete duality
if the dual Lagrangian $\bar{\cal L}(S)$ is the same function of $S$ as 
${\cal L}$ is of $F$.

%%%%%%%%%%%%%%%%%%%%%%%%%%%%%%%%%%%%%%%%%%%%%%%%%%%%%%%%%%%%%%%
\section{Two-potential Formulation}
%%%%%%%%%%%%%%%%%%%%%%%%%%%%%%%%%%%%%%%%%%%%%%%%%%%%%%%%%%%%%%%
\setcounter{equation}{0}

There exists a more symmetric Hamiltonian formulation, in which
electric and magnetic sources are treated on the same footing. It was given
in \cite{DGHT} for Maxwell theory but is also valid in $D=4k$. The
two-potential formulation for $D\neq 2(p+1)$ is not particularly useful.

\subsection{1--forms in D=4 }
%%%%%%%%%%%%%%%%%%%%%%%%%%%%%%%%%%%%%%%%

We start from the covariant action (\ref{action}) and perform the standard
Legendre transformation on the spatial components of the
vector potential $A_i$ to reach first order form.  Denoting 
by $\pi^i$ the momenta conjugate to $A_i$, we get
\be
I= \int d^4x [\pi^i \dot{A}_i +\pi^i \,  ^{*}\! G_{0i} - {\cal H}(\pi,B)
+ A_0(\pi^i,_i + j^0)] + I'_{C} + I_P
\label{int1}
\ee
with
\be
I'_{C} = \sum_n e_n \int A_k(z_n) dz^k_n.
\ee
In (\ref{int1}), we have set
\be
{\cal H}(\pi,B) = \pi^i F_{0i} - {\cal L}(F),
\ee
where $F_{0i}$ is expressed in terms of $\pi^i$
through $\pi^i = - H^{0i}$.  We then solve the 
Gauss law $\pi^i,_i + j^0 =0$ by using the
strings attached to the particles\cite{DHT}, writing the momentum $\pi$ 
as the curl of a second vector potential plus a string contribution
\be
- \pi^i =  \epsilon^{ijk} \partial_j Z_k + \sum_n e_n \int dy^0_n \wedge
dy^i_n \delta^{4}(x-y_n).
\label{pi}
\ee
The  minus sign choice establishes symmetry between electric 
($-\mbox{\boldmath $\pi$}$)  and magnetic fields.

The action then reduces to (\ref{int1}) without the $A_0$-term;
$\pi^i$ now stands for  the function
of the dynamical variables defined in (\ref{pi}).
We introduce the symmetric notation
\begin{eqnarray}
q^a_n &=& (g_n,e_n), \\
{\bf A}^a &=& ({\bf A}, {\bf Z}), \\
{\bf B}^a &=& ({\bf B}, - \mbox{\boldmath $\pi$}),
\end{eqnarray}
with $a$=1 (magnetic), 2 (electric); now
\be
{\bf B}^a =\mbox{\boldmath $\nabla$} \times {\bf A}^a + \mbox{\boldmath 
$\beta^a$},
\label{B}
\ee
with
\be
\mbox{\boldmath $\beta^a$ }= \sum_n q^a_n \: \int dy^0_n \wedge d{\bf y}_n \;
\delta^4
(x-y_n)
\label{beta}
\ee
and
\be
{\cal H} = {\cal H}({\bf B}^a).
\ee
If we also define
\be
\alpha^a_i = \frac{1}{2} \epsilon_{ijk} \sum_n q^a_n
\int dy^j_n \wedge dy^k_n
\label{alpha}
\ee
we get\footnote{The sum $\dot{\bf A}^a +  \mbox{\boldmath $\alpha^a$}$ was
called $E^a_i$ in \cite{DGHT}.  However, since
${\bf B}^2$  is the actual electric field (equal to $\dot{\bf A}^a +
\mbox{\boldmath $\alpha^a$}$ up to a gauge only on shell ), we shall not
use this notation here.}

\be
I[A^a, z_n, y_n] = \int d^4x [- {\bf B}^2 \cdot ({\bf \dot{A}}^1
+ \mbox{\boldmath $\alpha^1$}) - {\cal H}] + I'_{C} + I_P \;.
\label{int2}
\ee

To derive the manifestly symmetric formulation,
let us first notice the (symmetric) identity
\begin{equation}
\frac{1}{2} \int d^4x k_{ab} 
{\bf B}^a ({\bf \dot{A}}^b + \mbox{\boldmath 
$\alpha^b$})
-\frac{1}{2} \sum_n  k_{ab}
 q^a_n \int {\bf A}^b (z_n) \cdot d {\bf z}_n
= \int d^4 x \left( \partial_\mu V^\mu +\frac{1}{2}
\, k_{ab} \mbox{\boldmath $\alpha^a\beta^b$} \right) \ .
\label{identity}
\end{equation}
Here $k_{ab}$ is the {\it symmetric} matrix
\be
k_{ab} = \left( \begin{array}{ll}
0 & 1\\
1 & 0
\end{array}
\right)
.
\ee
The form of $V^{\mu}$ is easily worked out; the variational derivatives 
of the second term vanish (as in (\ref{total})), which implies that
\be
\frac{1}{2} \int d^4 x \, k_{ab} \mbox{\boldmath $\alpha^a\beta^b$}
= \int dt \; d\sigma \frac{\delta W}{\delta y^\alpha_n} \dot{y}^\alpha_n
+ \int dt \frac{\partial W}{\partial z^\alpha_n} \dot{z}^\alpha_n \; .
\label{total'}
\ee
Again, since the configuration space of the particles and the
strings is multiply connected because of the vetos, the
functional $W$, which exists locally, turns out to be  multiple-valued.

If one  adds to the action (\ref{int2}) the left-hand
side of (\ref{identity}), which has vanishing variation,
one obtains the manifestly symmetric form of \cite{DGHT}
\be
I = \int d^4 x \left[\frac{1}{2} \epsilon_{ab} {\bf B}^a \cdot
({\bf \dot{A}}^b + \mbox{\boldmath $\alpha^b$}) - {\cal H}\right]
+ I_C + I_P
\label{dualaction}
\ee
with
\be
I_{C} = \frac{1}{2}
\sum_n \epsilon_{ab}
q^b_n \int {\bf A}^a (z_n) \cdot d {\bf z}_n.
\label{dualityinvariant}
\ee
Note that the addition of  (\ref{identity}) to the action has had the
effect of bringing in a magnetic minimal coupling term,
on a par with the standard  electric minimal coupling
term, each with a factor $1/2$. [The importance
of this factor $1/2$ has been recently stressed in the
string context in \cite{Cheung}].

Since we have derived the action (\ref{dualaction}) from
the original Dirac action, they are equivalent.
The two-potential action was previously derived in \cite{SSen} for
sourcefree Maxwell theory
(see \cite{sao} for a further discussion of the equivalence of the
two-potential formulation and the Maxwell action in the
absence of sources).

It is of interest to rederive the quantization condition in
the symmetric formulation.  The constraints for the strings momenta
are now\be
\frac{\hbar}{i} \frac{\delta \Psi}{\delta y^i_n}
=  \frac{q^a}{2} \epsilon_{ijk} \epsilon_{ab} B^{(b)j} y'^k_n  \; \Psi
\label{dyon}
\ee
as follows from the kinetic term in the action
(\ref{dualityinvariant}) (only the kinetic term
$(1/2) \epsilon_{ab}{\bf B}^a ({\bf \dot{A}}^b + 
\mbox{\boldmath $\alpha$}^b)$
contributes to the momenta conjugate to the spatial
coordinates of the strings, since in the gauge $y^0_n = \tau_n$, the
time derivatives $\dot{y}^k_n$ occur only in $ \mbox{\boldmath $\alpha$}^b$ and not in $\mbox{\boldmath $\beta $}^b$).

The phase picked up by the wave function in the double-pass gets contributions
  $(1/2) \epsilon_{ab}q^a \bar{q}^b$ from
turning string 1 around $\bar{q}^b$, and $-(1/2) \epsilon_{ab}
\bar{q}^a q^b$ from turning string 2 around the dyon $q^b$.
There is a minus sign because the orientations of the two-surfaces
swept out by the strings in their rotation are opposite.
The two contributions are equal and add up
to $ \epsilon_{ab}q^a \bar{q}^b$, leading to
the duality-invariant quantization condition for dyons found
above (equation (\ref{eg}), which reads, in the symmetric notation
\be
\epsilon_{ab}q^a \bar{q}^b = n h.
\ee

Note again that the total phase for the double-pass is the same
as in the previous representations, as it should since the double-pass
is contractible.  The phases from the individual rotations are different,
however, reflecting the fact that different representations of the
fundamental group are involved.  In the symmetric description,
each rotation brings in half of the total phase.

\subsection{Duality Invariant Theories}
%%%%%%%%%%%%%%%%%%%%%%%%%%%%%%%%%%%%%%%%%%%%

The previous discussion including the quantization condition
holds independently of the
form of the Lagrangian ${\cal L(F_{\mu \nu}})$:
This is because the momenta conjugate to the string coordinates always 
take the same form, leading always to the same
quantum constraints (\ref{electric}), (\ref{magnetic}), (\ref{dyon}).

When the Lagrangian is  
duality invariant, the symmetric action (\ref{dualaction})
is manifestly invariant under $SO(2)$-duality
rotations.  This is because the kinetic term, which
involves the invariant tensor $\epsilon_{ab}$, is
duality invariant \cite{DGHT}.  The Hamiltonian is
also invariant, as it depends on the fields only
through the invariant combination
$\delta_{ab} {\bf B}^a \cdot {\bf B}^b$ and
($\epsilon_{ab}{\bf B}^a \times {\bf B}^b$) \cite{us}.

Contrary to what is sometimes asserted, duality invariance is
a standard symmetry in the sense that it
is not just a symmetry of the equations of motion but also a
symmetry of the action, with associated ``chiral" charge
\cite{BB}.  Its canonical generator is simply the Chern--Simons
spatial
integral\cite{DGHT,DHT}
\be
G = - \frac{1}{2} \int d^3x \, {\bf A}^a \cdot
{\bf B}^b \delta_{ab} \; .
\label{ChernSimons}
\ee
The invariance of the Hamiltonian is equivalent
to $[H, G]= 0$. [In the one-potential formulation \cite{DHT}, 
$G$ is the same, with ${\bf A}^2$ replaced by
$ \nabla^{-2}( \nabla \times\mbox{\boldmath $\pi$})$.]

\subsection{Two-potential Formulation in D=6}
%%%%%%%%%%%%%%%%%%%%%%%%%%%%%%%%%%%%%%%%%%%%%%%%%%%%%%%%

The symmetric magnetic/electric description proceeds as in D=4:
We  introduce the symmetric notation
\begin{eqnarray}
q^a_n &=& (g_n,e_n), \\
A^a_{ij} &=& (A_{ij}, Z_{ij}), \\
B^a_{ij} &=& (B_{ij}, E_{ij}),
\end{eqnarray}
again with $a$=1 (magnetic), 2 (electric).  Here
$B^{ij}$ is the magnetic field,
\be
B^{ij} = \frac{1}{3!} \epsilon^{ijklm} F_{klm},
\ee
and $Z_{ij}$ the potential for the electric field.  In this unified form,
\be
B^{a \, ij} = \frac{1}{2} \epsilon^{ijklm} \partial_k A_{mn}^a
- \beta^{a \, ij} \; ,
\ee
where the membrane contribution $\beta^{a \, ij}$ is
\be
\beta^{a \, ij} = \sum_n q^a_n \int d^6x dy^0_n \wedge dy^i_n
\wedge dy^j_n \delta^6(x-y_n)\; .
\ee
Similarly, we introduce
\be
\alpha^a_{ij} = \frac{1}{3!} \epsilon_{ijklm} \sum_n q^a_n \int d^6x
dy^k_n \wedge dy^l_n \wedge dy^m_n \delta^6(x-y_n) \; .
\ee

Repeating the steps that led to the action (\ref{int2}) and
observing that the momentum conjugate to $A_{ij}$ is now $(-1/2)
E^{ij}$, one finds for the action,
\be
I[A^a, z_n, y_n] = \int d^6x [- {\bf B}^2 \cdot ({\bf \dot{A}}^1
+\mbox{\boldmath $\alpha^1$} )  - {\cal H}] + I'_{C} + I_P
\label{int2bis}
\ee
where ${\cal H}(B^a_{ij})$ is the $2$-form Hamiltonian
density, introducing the obvious convention
\be
{\bf N} \cdot {\bf M} \equiv \frac{1}{2} N^{ij} M_{ij} \; .
\ee
In (\ref{int2bis}), $I'_{C}$ is
the spatial part of the electric minimal coupling term.
If (as before) one now adds the integral of a suitable total divergence,
the action becomes symmetrized:
\be
I = \int d^6 x [- \frac{1}{2} k_{ab} {\bf B}^a \cdot
({\bf \dot{A}}^b + \mbox{\boldmath $\alpha^b$}) - {\cal H}]
+ I_{C} + I_P
\label{dualactionbis}
\ee
\be
I_{C}\equiv  \frac{1}{4}
\sum_n k_{ab}
q^b_n \int A_{ij}^a (z_n) dz^i_n \wedge dz^j_n \; .
\label{dualityinvariantbis}
\ee
The one difference from the 1-form case is simple but far-reaching: 
$\epsilon_{ab}$ is replaced by  $k_{ab}$.
Its symmetry ensures that  of the kinetic term under exchange
of ${\bf A}$ and ${\bf Z}$, due to the symmetry of $\epsilon^{ijklm}$
under the permutation of any two pairs of indices.
As pointed out in \cite{DGHT}, the fact that
it is the non-duality invariant $k_{ab}$ that appears
destroys the invariance of the kinetic term -- and hence also of the
action -- under $SO(2)$-rotations of the
$B^a_{ij}$, even when the Hamiltonian is invariant.

\subsection{Charge Quantization}
%%%%%%%%%%%%%%%%%%%%%%%%%%%%%%%%%%%%

As a consequence of the modification of the kinetic
term,  the quantum constraints in the symmetric formulation
read
\be
\frac{\hbar}{i}\frac{\delta \Psi}{\delta y^k_n} =
- \frac{q^a_n} {4} k_{ab}
E^{b \, ij} \epsilon_{ijklm}  \frac{\partial y^l_n}
{\partial \sigma^1_n}  \frac{\partial y^m_n}{\partial \sigma^2_n} \Psi  \ .
\ee
Thus, the phase
acquired in the double pass through the motion of
the first membrane is $(1/2 \hbar) k_{ab} q^a \bar{q} ^b$,
while the phase acquired through the motion of the second membrane
is again $(1/2 \hbar) k_{ab} q^a \bar{q} ^b$.
Demanding that the total phase is a multiple of $2   \pi$
leads then to the symmetric quantization condition
\be
k_{ab} q^a \bar{q} ^b = n h .
\label{plusk}
\ee
Note that this treatment gives a   way to check the orientations. Had they
been
opposite one would have obtained zero, in contradiction with what one obtains
for single electric and magnetic poles, which are particular cases of dyons.

The condition (\ref{plusk}) is not empty for a single dyon
and there reads 
\be
k_{ab} q^a q^b = 2 e g = n h.
\ee
For a self-dual source ($e = g$), one gets $2 (e)^2 = n h$,
as in \cite{DGHT} (the factor $2$ is absent there because of
different normalization conventions).

\subsection{Chiral/Anti-chiral Decomposition}
%%%%%%%%%%%%%%%%%%%%%%%%%%%%%%%%%%%%%%%%%%%%%%%%%%%%%%%%%%%%%%%%%%

We saw that $SO(2)$-rotations of the fields are not
invariances of the action here,  due to the
non-invariance of $k_{ab }$ in the kinetic term: Duality rotations are
not even canonical transformations \cite{DGHT}.  There can only be a residual
discrete $Z_2$-symmetry,
\begin{eqnarray}
{\bf E} \rightarrow {\bf B} &,& {\bf B} \rightarrow {\bf E}
\\
e  \rightarrow g &,& g \rightarrow e
\end{eqnarray}
under which the quantization condition is clearly invariant
\cite{DGHT}.  This motivates a decomposition  into chiral and anti-chiral
components, defined through
\begin{equation}
A_{ij} \pm Z_{ij} = \sqrt{2} \, U^{\pm}_{ij} ,
\end{equation}
and consequently also
\begin{equation}
B_{ij} \pm E_{ij} = \sqrt{2} \, V^{\pm}_{ij} ,
\end{equation}
where $V^{\pm}_{ij}$ are respectively defined in terms of
$U^{\pm}_{ij}$ as ($B_{ij}, E_{ij}$) are defined in terms of
($A_{ij}, Z_{ij}$) 
\begin{equation}
V^{\pm}_{ij} = \frac{1}{2} \epsilon^{ijklm} \partial_k U^{\pm}_{mn}
-\frac{1}{\sqrt{2}} [\beta^{1 \, ij} \pm \beta^{2 \, ij}] \; .
\end{equation}
The factor $\sqrt{2}$ has been inserted to match the
normalization conventions of \cite{DGHT} for the kinetic term.
We also define $e_\pm = (g \pm e)/\sqrt{2}$.
If $g = e$ (self-dual source)
only $e_+ \not= 0$; if $g = -e$ (anti-self-dual source),
only $e_- \not= 0$.  Note that for anti-self-dual sources,
$e_- = \sqrt{2} e$ is the strength of the anti-self-dual source used in
\cite{DGHT} (and called there self-dual).

The kinetic term of  the symmetric action (\ref{dualactionbis}) splits into the sum of two
non-interacting pieces.   In the free case  this is also true of the Hamiltonian,  and one gets, 
for the anti-self-dual part,
\be
I^- = \frac{1}{4} \int d^6 x [ {\bf V^{-}} \cdot
({\bf \dot{U}^{-} } +\mbox{\boldmath $\alpha_{-}$} - {\bf V^{-}}) ]
+ I_{C} + I_P
\ee
which is the action (22) of  \cite{DGHT}.
Here,
\be
I_{C}
= - \frac{1}{4} \int d^6x A_{ij} J^{ij} , \;\;\;\; J^{ij} = \sum_n e_{n} \int
dz^i_n \wedge dz^j_n \delta^6(x-z_n) \; .
\ee

The quantization condition becomes
\be
\eta_{ab} e^a \bar{e}^b = n h
\label{egDual}
\ee
with
\be
\eta_{ab} = \left( \begin{array}{ll}
1 & 0\\
0 & -1
\end{array}
\right)
\ee
and $e^a = (e_+,e_-)$.
This implies, for  purely self-dual (anti-self-dual) sources, that
$e_+ \bar{e}_+ = nh$ ($e_-\bar{e}_- = nh$ ).  Furthermore,
there is no condition on the relative strengths of
self-dual and anti-self-dual pure sources (if there are only
pure sources).  Conversely,
one can reassemble the quantization for the individual self-dual
or anti-self dual pieces obtained in
\cite{DGHT} to get (\ref{egDual}).

%%%%%%%%%%%%%%%%%%%%%%%%%%%%%%%%%%%%%%%%%%%%%%%%%%%%%%%%%%%%%%%%%%%%%%%%%%%%%

The extension  of our results
to higher dimensions is a straightforward addition of indices.
For $D=4k$, $p=2k-1$, one follows the pattern of ordinary electrodynamics
(Section 2), whereas for $D=4k+2$, $p=2k$ (including 0 forms in
D=2 \cite{HT}), one follows the discussion of section
3 ($D=6,\ p=2$).   The results remain unchanged.

\setcounter{equation}{0}

{\it Note added:}~ While this paper was being completed, we received the 
preprint \cite{244}, in which the plus sign in the quantization
condition for dyons in $4k+2$ dimensions is also stressed.
Interestingly it is derived there by embedding the theory in a
supergravity model with a continuous, rigid  $SO(5,5)$,
duality invariance group \cite{Cremmeretal}.  Acting with this group on
a system of purely electric and purely magnetic
membranes, for which the quantization condition
was derived in  \cite{CT,Nepo}, one
obtains dyons which fulfill $e \bar{g} + \bar{e} g = nh$.
Our derivation shows that this quantization condition is actually
independent both of any particular embedding of the abelian
$(2k)$-form theory in a bigger model, and also of the existence
of a rigid duality invariance group for the embedding theory.

\section*{Acknowledgements}
%%%%%%%%%%%%%%%%%%%%%%%%%%%%%%%%%

Discussions with Frank Wilczek are gratefully acknowledged.
The work of SD was supported by the National Science Foundation,
grant \#PHY-9315811, that of MH was partly supported
by  research funds from IISN (Belgium),
that of AG and CT by Grants 3960008 and 1970151
of FONDECYT (Chile).  AG and
CT also acknowledge institutional
support to the
Centro de Estudios Cient\'{\i}ficos de Santiago provided by
SAREC (Sweden) and a group of Chilean private companies
(BUSINESS DESIGN ASSOCIATES, CGE, CMPC, CODELCO, COPEC, 
CHILGENER, MI\-NE\-RA COLLAHUASI, MI\-NE\-RA LA ESCONDIDA,
 NOVAGAS, XEROX CHILE).
Finally, AG would like to gratefully acknowledge the hospitality of the
Institute for Advanced Study while this work was being completed. The clerical
assistance  of Santenber Belices is acknowledged.

%%%%%%%%%%%%%%%%%%%%%%%%%%%%%%%%%%%%%%%%%%%%%%%%%%%%%%%%%%%%%%%%%%%%%%%%%%%%%
%%%%%%%%%%%%%%%%%%%%%%%%%%%%%%%%%%%%%%%%%%%%%%%%%%%%%%%%%%%%%%%%%%%%%%%%%%%%%
%%%%%%%%%%%%%%%%%%%%%%%%%%%%%%%%%%%%%%%%%%%%%%%%%%%%%%%%%%%%%%%%%%%%%%%%%%%%%
\appendix
%%%%%%%%%%%%%%%%%%%%%%%%%%%%%%%%%%%%%%%%%%%%%%%%%%%%%%%%%%%%%%%%%%%%%%%%%%%%%

%%%%%%%%%%%%%%%%%%%%%%%%%%%%%%%%%%%%%%%%%%%%%%%%%%%%%%%%
\section*{Appendix A. Geometry  of the ``double pass''}
%%%%%%%%%%%%%%%%%%%%%%%%%%%%%%%%%%%%%%%%%%%%%%%%%%%%%%%%
\renewcommand{\theequation}{A.\arabic{equation}}
\setcounter{equation}{0}

We first perform the analysis in three space dimensions and then extend it to
higher dimensions.

Consider the situation illustrated in Fig. 1, as described in the caption.
At each instant of the turning, 
we may imagine each dyon moving by
``eating up'' its own string,  so that they both end up, again on the $x^3$
axis, but with dyon 2 on the left of dyon 1. Since the two strings never touch
while the double pass is generated, this displacement of the dyons does not
violate the Dirac veto. The operation deforms the whole surface generated by 
the double pass into two single strings, which correspond to a single point in
configuration space.  Note that in the original formulation of Dirac, 
where strings are only attached
to the magnetic poles, the same procedure shows that the ``single pass'' used 
by
Dirac is contractible to a point also. However, when strings are attached to
both charges, the single pass (in which only one string turns) is not
contractible because one would violate the Dirac veto: at the instant when the
two strings cross, the displacement needed to flip the dyons in the above
argument would make one dyon go through the string of the other.

The displacement of string 1 around dyon 2 generates a surface with the
topology of a two--sphere, which bounds a three--ball with
positive orientation relative to
\be
dx^1\wedge dx^2 \wedge dx^3
\label{a1}\ \ ,
\ee
whereas the orientation associated with the displacement
around dyon 1 is the same as 
\be
dx^1\wedge dx^2 \wedge (-dx^3) = - dx^1\wedge dx^2 \wedge dx^3 \ \ .
\label{a2}
\ee
The  $-dx^3$ in  (\ref{a2}) instead of the $+dx^3$ in
(\ref{a1})  reflects the Dirac strings going from 2 to 1 being oriented
opposite to
that going from 1 to 2. The turning that generates the double pass takes place
in the $(x^1,x^2)$ plane, and since both strings turn in the same
direction, we
have written $+dx^1\wedge dx^2$ in both cases. Comparison of (\ref{a1}) and
(\ref{a2}) shows that the two spheres have opposite orientation, giving
rise to
the antisymmetric quantization rule described in section 2.2 of the main text.

We now proceed to higher dimensions.  Consider first, for definiteness, the
case of a 
two--form potential in D=6
spacetime as discussed in section 3. The dyons are then
infinite strings in D=5  space. We take them to lie
along the 4- and 5-axes respectively, separated by a distance $a$ along the
3-axis  (a configuration that will be generalized  in Appendix C).

In the double pass, the membrane of dyon 1 generates a $3$--sphere which links
dyon 2 and vice versa. Without loss of generality one may take the $3$--sphere
linking the first dyon to lie in the subspace $x^4=0$ and that linking the
second dyon to lie in the subspace $x^5=0$. The rotation that generates the
double pass takes place in the $(x^1,x^2)$ plane, which is complementary to
the
subspace containing the two dyons and the line that joins them, just as 
in three space dimensions.

We are now interested in the orientation of these two $3$--spheres relative to
the orientation of the five dimensional space and to that of the dyons
themselves (in the three dimensional case the dyons were points and had only 
one intrinsic orientation).  As in the three dimensional case, 
the orientation of the sphere linking dyon 2 may be taken to be positive
relative to
\be
dx^1 \wedge dx^2 \wedge dx^3 \wedge dx^4 \wedge dx^5  \ \ ,
\label{a3}
\ee
whereas for the sphere linking dyon 1 we have
\be
dx^1 \wedge dx^2 \wedge (-dx^3) \wedge dx^5 \wedge dx^4 =
+ dx^1 \wedge dx^2 \wedge dx^3 \wedge dx^4 \wedge dx^5 \ \ .
\label{a4}
\ee

Also as before, we have replaced $dx^3$ in (\ref{a3}) by $-dx^3$  
in (\ref{a4}), because exchanging the dyons reverses the orientation of 
the coordinate along the line that joins them and we have taken
$dx^1\wedge dx^2$ with the same sign in both cases by definition of the double
pass. There is now, however, a new feature, which is the exchange of the $x^4$
and $x^5$ coming from the exchange of the dyons. Due to this, the surfaces
have
the same orientation, giving rise to the symmetric quantization indicated
in text.

The extension to a $p$--form potential in $2p+1$ space dimensions is
immediate.
Each dyon contributes a $(p-1)$--form, $\omega^{(p-1)}$, to the exterior
product, so that one has
\be
dx^1\wedge dx^2 \wedge \ldots \wedge dx^{(2p+1)}= dx^1\wedge dx^2 \wedge
dx^3 \wedge \omega^{(p-1)}_{1} \wedge \omega^{(p-1)}_{2}  \ .
\label{a5}
\ee
Under exchange of the two dyons on the double pass, one has
\begin{eqnarray}
dx^1\wedge dx^2 &\longrightarrow& dx^1 \wedge dx^2 \\
dx^3 &\longrightarrow& -dx^3 \\
\omega^{(p-1)}_{1} \wedge \omega^{(p-1)}_{2}  &\longrightarrow&
(-1)^{(p-1)} \omega^{(p-1)}_{1} \wedge \omega^{(p-1)}_{2} \ \ .
\end{eqnarray}
Hence the relative orientation of the two surfaces is $(-1)^p$, giving rise
to antisymmetric quantization for odd $p$ and symmetric quantization for even
$p$.

%%%%%%%%%%%%%%%%%%%%%%%%%%%%%%%%%%%%%%%%%%%%%%%%%%%%%%%%
\section*{Appendix B. Quantizations for Even p}
%%%%%%%%%%%%%%%%%%%%%%%%%%%%%%%%%%%%%%%%%%%%%%%%%%%%%%%%

\renewcommand{\theequation}{B.\arabic{equation}}
\setcounter{equation}{0}

It was observed long ago that the quantization condition
in four dimensions does not force the electric charge carried by
a dyon to be an integer multiple of the minimum electric charge 
$e_0$ carried by
purely electric particles \cite{Schwinger,Zwan,Olive}.  However, if one
imposes CP conservation, the electric charge carried by dyons must be
an integer, or  half-integer, multiple of $e_0$ \cite{Witten}.  We shall
show that the same result holds also for $2p$-forms, but without
having to invoke CP conservation explicitly.

The general solution of the quantization condition
\be
k_{ab} q^a \bar{q} ^b = e \bar{g} + g \bar{e}  = n h
\ee
can be presented in the chiral-anti-chiral basis, in which
(\ref{egDual}) shows that the allowed values of the charges form a
Lorentzian lattice.  However, for comparison with the $2p +1$-case,
we shall discuss the quantization condition in its 
original form and shall assume that there are purely
electric sources.

Let $e_0$ be the minimum value for the electric charge.  Then,
the quantization condition shows that the magnetic charge
of dyons is an integer multiple of $(2 \pi \hbar)/e_0$,
\be
e_0 g = 2 \pi n \hbar.
\ee
Consider a dyon with the minimum allowed magnetic charge
$g_0 = (2 \pi \hbar)/e_0$.
Since the quantization condition for a single dyon is non-empty,
and reads $2 e g = 2 \pi m \hbar$,  the
electric charge $e$ carried by this dyon should fulfill
\be
2 e = m e_0,
\label{elementarycharge}
\ee
as announced.  

If the integer $m$ in (\ref{elementarycharge}) is even, then
the electric charge carried by any dyon must be an integer multiple
of $e_0$.  This follows from the quantization applied
to the above ``reference" dyon and any other dyon.
The charges of a general dyon are thus given by
$(k e_0, l g_0)$ with $k$, $l$ integers. 

If, on the other hand, the integer $m$ is odd, so that the
above dyon has an electric charge which is a half-integer multiple
of $e_0$, then, all dyons with odd magnetic charge have also an
electric charge that is a half-integer multiple of $e_0$, while
all dyons with even magnetic charge have an electric charge which
is an integer multiple of $e_0$.

In D=4, one may shift the electric charge of dyons by
adding a CP-violating term $\theta\int F \wedge F$ to the action
\cite{Witten}.  This
possibility does not exist in D=4 $p +2$ because the
curvature form $F$ is then of odd degree, so that $F \wedge F$ is
identically zero.

%%%%%%%%%%%%%%%%%%%%%%%%%%%%%%%%%%%%%
\section*{Appendix C. Quantization From Angular Momentum}
%%%%%%%%%%%%%%%%%%%%%%%%%%%%%%%%%%%%%

\renewcommand{\theequation}{C.\arabic{equation}}
\setcounter{equation}{0}

The quantization condition for the 
product of electric and magnetic charges
has a very appealing physical basis \cite{Fierz} in terms of
the spectrum of the angular momentum operator.
We show here how this argument extends to dyons in higher dimensions
and especially how it brings in the two different types of quantization
for even or odd $p$ in a natural way.  Recall that in D=2($p$+1), the electric
and magnetic fields are spatial $p$-forms and their sources are
($p$-1) spatial extended objects.

We begin by revisiting the original argument in D=4. 
Consider a magnetic pole of strength $g$ located at the origin of
coordinates and an electric pole of strength $e$ at the point $(0,0,a)$. The
total angular momentum stored by the field points in the $x_3$ direction,
and is
\begin{eqnarray}
J_{12} &=&\int d^3x (x_1T_{02}-x_2 T_{01}) = \int \left[{\bf r}\times({\bf
E}\times{\bf B})\right]_3 d^3 x \nonumber \\
&=& -2eg \int d^3 x \Phi^{E}\frac{\partial}{\partial x_3} \Phi^{M} 
=\frac{eg}{4\pi} \ \ ,
\label{j}
\end{eqnarray}
in terms of the corresponding scalar potentials
\begin{eqnarray}
\Phi^{E} &=& \frac{1}{4\pi(x_1^2+x_2^2+(x_3-a)^2)^{1/2}} \label{scalare} \\ 
\Phi^{M} &=& \frac{1}{4\pi(x_1^2+x_2^2+x_3^2)^{1/2}} \label{scalarm} \ .
\end{eqnarray} 
The half-integer quantization of this angular momentum
yields the Dirac quantization condition.

Note that the  integral in (\ref{j}) does not depend on the parameter $a$ 
as long as this parameter is different from zero [this can be seen before 
doing the integral just from dimensional analysis]; for $a$=0 the integral
manifestly vanishes. Thus there is a striking difference 
between an electric pole and
a magnetic pole which are very close to each other and a dyon where the two
poles are at the same point.

The angular momentum is  antisymmetric under the exchange of the two poles,
which amounts to exchange of {\bf E} and {\bf B};
therefore if we have two dyons, the angular momentum will be given by
\begin{equation}
J_{12}= \frac{1}{4\pi}(\bar{e}g-e\bar{g}) \ \ ,
\end{equation}
for  dyons of charge $(e,g)$ and $(\bar{e},\bar{g})$, located at $(0,0,0)$ and
$(0,0,a)$ respectively. This gives the quantization condition (\ref{minus}).

Consider now the general  case in $D=2(p+1)$ dimensions. An electric pole,
which is now a $(p-1)$--dimensional extended object,  is located at
$(0,0,a;x_{a_1},\ldots,x_{a_{p-1}}; 0 , \ldots,0)$ , with $-\infty < x_a <
\infty$. At the same time, a magnetic pole is located at $(0,0,0; 0 ,
\ldots,0;x_{b_1},\ldots,x_{b_{p-1}})$, with $-\infty < x_b < \infty$, $4\le
a \le p+2$, $p+3\le b \le 2p+1$.

The electric--magnetic  fields produced by these poles 
are derived from two potentials  
$\Phi^E$--$\Phi^M$,   which are spatial $p-1$ forms that generalize the
scalar potentials (\ref{scalare}) and  (\ref{scalarm}) for $p=1$,
\begin{eqnarray}
H^0_{\ i_1\ldots i_p} &=&(d\Phi^E)_{i_1\ldots i_{p-1},i_p}  \ ,\\
^* \! F^0_{\ i_1\ldots i_p} &=&(d\Phi^M)_{i_1\ldots i_{p-1},i_p} \ , \\
\end{eqnarray}
and whose only nonvanishing component are
\begin{eqnarray}
\Phi^E_{a_1\ldots a_{p-1}} &=& 
-\frac{e\epsilon_{a_1,\ldots a_{p-1}}}{p S_{p+1}(x_1^2+x_2^2 +
(x_3-a)^2+x_{b_1}^2+\cdots +x_{b_{p-1}}^2)^{\frac{p}{2}}} \ \label{phie} ,\\
\Phi^M_{b_1\ldots b_{p-1}} &=& 
-\frac{g\epsilon_{b_1,\ldots b_{p-1}}}{p S_{p+1}(x_1^2+x_2^2 +
x_3^2+x_{a_1}^2+\cdots+ x_{a_{p-1}}^2)^{\frac{p}{2}}} \label{phim} \ ,
\end{eqnarray}
where 
\begin{equation}
S_{p+1} = \frac{2\pi^{\frac{p+2}{2}}}{\Gamma\left(\frac{p+2}{2}\right)}
\label{sphere}
\end{equation} 
is the area of the $(p+1)$--sphere.

The angular momentum stored in the field is given by
\begin{equation}
J_{ij} = \int d^{D-1} x ( x_i T_{0j} - x_j T_{0i} ) \ ,
\label{am}
\end{equation}

where  $T_{0i}$ is given by 
\begin{equation}
T_{0i} = \frac{1}{(p!)^2}\epsilon_{i j_1\cdots j_p \ k_1 \cdots k_p} H^{0
j_1\cdots j_p}
{}^*\! F^{0k_1 \cdots k_p} \ .
\label{t0i}
\end{equation}
It is convenient to work with the spatial dual , $^\dagger \! J$ of
the angular momentum, which in this case is 
\begin{eqnarray}
^\dagger \! J_{i_1 \ldots i_{2p-1}} &=& \frac{1}{2} \epsilon_{i_1 \ldots
i_{2p-1}lm}\int d^{D-1} x J_{lm} 
 \label{int}\\
&=& \frac{1}{(p-1)!^2}\delta^{[j_1\cdots j_p \ k_1 \cdots k_p]}_{[i_1
\cdots \cdots\cdots i_{2p-1}l]}
\int d^{D-1} x \ x^l \Phi^{E}_{j_1\cdots j_{p-1},j_p}\Phi^{M}_{k_1\cdots
k_{p-1},k_p} \nonumber \\
&=& -\frac{2(-1)^p}{(p-1)!^2}\delta^{[j_1\cdots j_{p-1} \ k_1 \cdots
k_p]}_{[i_1  \cdots \cdots\cdots\cdots \ i_{2p-1}]}
\int d^{D-1} x \  \Phi^{E}_{j_1\cdots j_{p-1}}\Phi^{M}_{k_1\cdots k_{p-1},k_p}
\ , \nonumber 
\end{eqnarray}
where $\delta^{[123\ldots]}_{[123\ldots]}=1$ and we have performed an
integration by parts, dropping  a vanishing contribution at infinity.
It is
easy to see from the symmetries of the integral in (\ref{int}) that only 
\begin{eqnarray}
&&J_{12} =  ^\dagger \! J_{3\ 4\ \cdots \ (2p+1)} =  -2(-1)^p\int d^{D-1} x \
\Phi^E_{4\cdots (p+2)}
\Phi^E_{(p+3)\cdots (2p+1),3}  \label{j12}\\  
&=& -\frac{(-1)^p eg\Gamma(\frac{p}{2})^2}{8\pi^{p+2}}
\int  \ \frac{d^{D-1} x}{(x_1^2+x_2^2+(x_3-a)^2+x^b x_b)^{p/2}}
\frac{\partial}{\partial x_3}\left(\frac{1}{(x_1^2+x_2^2+x_3^2+x^a
x_a)^{p/2}}\right) 
\nonumber
\end{eqnarray}
survives.  The integral in (\ref{j12}) has the key property that upon
integration over one of the $x^a$ and one of the $x^b$ coordinates it
yields (up to a sign, which depends on the choice of orientation), the same
expression with $p$  replaced by $p-1$. This follows from
\begin{equation}
\int_{-\infty}^{\infty} \frac{dx}{(a + x^2)^{p/2}} =
\frac{\sqrt{\pi}\Gamma(\frac{p-1}{2})}{a^{\frac{p-1}{2}}\Gamma(\frac{p}{2})}
 \ .
\end{equation}  
Hence the result for the calculation of the angular momentum of an electric
pole of charge $e$ and a magnetic pole of charge $g$ is (up to a sign) the
same for all dimensions.

The symmetry of the angular momentum under the exchange of the two poles is
now $(-1)^p$ as can be seen from (\ref{int}). This shows that the
quantization condition for dyons
is symmetric for even $p$ and antisymmetric for odd $p$.

It is of interest to point out that again the integral (\ref{int}) is independent of  the distance $a$ between  the poles as long as $a$ is different from zero.  The integral  vanishes for $a=0$ and also for $a\neq 0$ if the  two $(p-1)$--dimensional sources and the vector $a^{i}$ define a subspace whose dimension is less than   $(2p-1)$.  For the angular momentum not to vanish we need that subspace to be of maximal dimension $(2p-1)$. Then the angular momentum  lies in its orthogonal plane.

%%%%%%%%%%%%%%%%%%%%%%%%%%%%%%%%%%%%%
\section*{Appendix D. Local gauge charts}
%%%%%%%%%%%%%%%%%%%%%%%%%%%%%%%%%%%%%
\renewcommand{\theequation}{D.\arabic{equation}}
\renewcommand{\thesubsection}{D.\arabic{subsection}}
\setcounter{equation}{0}

One may also derive the quantization conditions for $p$--brane dyons using
the 
analysis based on twisted connections and gauge patches a la Wu and 
Yang\cite{WY}. This may be done both in the two--potential and one--potential 
formulations. We will carry out the discussion for  $p=1$, $D=4$. The 
generalization to higher dimensions is straightforward\cite{HT2}.

\subsection{Two--potential formulation}
%%%%%%%%%%%%%%%%%%%%%%%%%%%%%%%%%%%%%%%

The key point is to realize that the connection $\omega$ is given by
\begin{equation}
\omega_{j}= \frac{i}{\hbar} \epsilon_{ab} q^{b} A^{a}_{j} \ ,
\label{connection}
\end{equation}
without the factor 1/2 one might naively expect from the explicit interaction 
term
\be
I_{int.} = \frac{1}{2}
\epsilon_{ab}
q^b \int {\bf A}^a \cdot d {\bf z}
\label{dualityinvariant2}
\ee
appearing in (\ref{dualityinvariant}).

To derive the connection one needs to consider a single particle and relate 
the momentum $p_{i}$ canonically conjugate to its position $z^i$, to the 
mechanical momentum
\be
P_{i} = \frac{m\dot{\bf z}}{\sqrt{1-\dot{\bf z}^2}}
\ee
(for simplicity we parameterize $y^0=\tau$ for the worldsheet).
This is because $ip_j/\hbar$ corresponds 
to the partial derivative $\partial/\partial z^j$ 
whereas $iP_j/\hbar$ corresponds to the covariant derivative
\be
\nabla_{i} = \frac{\partial}{\partial z^i} + \omega_i \; .
\label{covariantd}
\ee
Therefore we have 
\be
\frac{\hbar}{i}\omega_j = P_j  - p_j \ .
\ee
Now, to evaluate the canonical momentum we need the part of the action which 
contains time derivatives. That part is given by the sum of 
(\ref{dualityinvariant2}) and the contribution
\be
\int d^4 x \frac{1}{2} \epsilon_{ab}(\mbox{\boldmath $\alpha^b$}\cdot {\bf
B^a} - 
\mbox{\boldmath $\beta^b$}\cdot \partial_0 {\bf A^a}) \ ,
\ee
which, by using (\ref{beta}), (\ref{alpha}) and performing the integral over 
$d^4 x$ may be written as 
\begin{eqnarray}
&&\frac{q^b}{2}\epsilon_{ab} \int\left(\partial_j A^a_i dy^j \wedge dy^i + 
\partial_0 A^a_{i} dy^0 \wedge dy^i \right) \nonumber \\
&=& \frac{q^b}{2}\epsilon_{ab} \int d[A_i(y) dy^i] \nonumber \\
&=& \left.\frac{q^b}{2}\epsilon_{ab}\int_{\tau_1}^{\tau_2}  A^a_i(z) dz^i  +
 \frac{q^b}{2}\epsilon_{ab} \int A^a_i(y) dy^i  \right|^{\tau_2}_{\tau_1} \ .
\end{eqnarray}
This last form  combined with (\ref{dualityinvariant2}) shows that the 
connection is indeed given by (\ref{connection}).

The connection produced by a dyon of charge $Q^a$ at the origin may be taken 
to have only non vanishing spherical component
\be
A^{a}_{\phi} = \frac{Q^a}{4\pi}(1\pm \cos \theta) \ .
\label{dyonA}
\ee
In (\ref{dyonA}) the upper sign excludes the North pole $\theta=0$, whereas 
the lower sign excludes the South pole $\theta=\pi$. Demanding that the two 
gauge patches overlap so as to give the same transport along -- say -- the 
equator yields the desired quantization condition
\be
2\pi i n = \int (\omega_{\phi}^{+} - \omega_{\phi}^{-} )d\phi = 
\int \frac{i}{2 \pi \hbar} \epsilon_{ab} q^b Q^a d\phi = 
\frac{i}{\hbar}\epsilon_{ab}Q^a q^b \ .
\ee

\subsection{One--potential formulation}
%%%%%%%%%%%%%%%%%%%%%%%%%%%%%%%%%%%%

The one-potential formulation given in the text can also accommodate dyons but it 
only exhibits a connection for transporting either purely
electric charges or 
purely magnetic ones. In the first case the connection is the usual
$A_{\mu}$, 
whereas in the second it is the $Z_{\mu}$ introduced in Sec. \ref{dual4}.

However, in either case the connection, which does not appear explicitly
is also present, as can be seen from the following argument.  We will work
for definiteness in the usual formulation with $A_{\mu}$. 
The Lagrangian depends on the velocity $\dot{z}^{\mu}$ of each particle
both explicitly through  the kinetic and minimal coupling terms, and also
implicitly as  the velocity of the end point of the attached Dirac string.
It is through this last dependence that the ''missing'' connection
$Z_{\mu}$ enters, much in the same way as the ``missing'' $1/2$ in the
two--potential formulation is found. 

Indeed, the Dirac string velocities enter the action through the term
\begin{equation}
-g\int {}^*\! H = -g\int {}^*\! H_{\mu\nu}\frac{\partial y^{\mu}}{\partial
\tau}
\frac{\partial y^{\nu}}{\partial \sigma} d\tau d\sigma \ ,
\label{aa}
\end{equation} 
as is most easily seen by inserting $\pi_{\mu}$ obtained from the covariant
form 
\begin{equation}
\pi_{\mu} = -g {}^*\! H_{\mu\nu}\frac{\partial y^\nu}{\partial\sigma}\ ,
\label{covpi}
\end{equation}
of the constraint (\ref{mom}) into the kinetic term
$\int\pi_{\mu}\dot{y}^{\mu}d\sigma d\tau$ of the Hamiltonian action.

Now, in the Wu--Yang analysis, $^*\! H$ in (\ref{aa}) is the field produced
by a fixed source dyon of charge $(\bar{e},\bar{g})$ located, say, at the
origin of coordinates. Thus one may write
\begin{equation}
^*\! H = dZ + ^* \! M \ ,
\label{cc}
\end{equation}
where $Z$ is the dual connection and $^*\! M$ is the Dirac string
contribution of the source dyon (see Sec. 2.3). 

Next, one  argues that  by choosing the Dirac string of the source dyon not
to cross the Dirac string of the test dyon  $(e,g)$, $M$ may be taken equal
to zero in the integral (\ref{aa}). This is not only possible, but also
mandatory if one wants to have a non--singular $Z_{\mu}$ as is essential
for the Wu--Yang analysis, which uses regular connections over local
patches ($^*H$ is assumed to be regular away from its source, so if $M$ is
different from zero, $dZ$ must be singular and hence so must be $Z$ ). 

Once $M$ is set equal to zero, one may convert the surface integral
(\ref{aa}) into a line integral over the boundary of the worldsheet by
means of the Stokes theorem to obtain
\begin{eqnarray}
-g\int {}^*\! H &=& -g\oint Z  \nonumber \\ 
&=& -g \int_{\tau_{1}}^{\tau_{2}} Z_{\mu}\frac{\partial z^{\mu}}{\partial
\tau} d\tau
-g \int_{\sigma=0}^{\infty} Z_{\mu}\frac{\partial y^{\mu}}{\partial \sigma}
d\sigma \; .
\end{eqnarray}
This last relation shows that the Dirac string yields a
contribution$-gZ_{\mu}$ to the conjugate momentum of the particle.
Therefore the total connection is
\be
\omega_{\mu} = eA_{\mu} - gZ_{\mu} \ .
\label{con}
\ee

Note that the minus sign in front of the magnetic charge agrees with that
of the minimal coupling term in Eq. (\ref{min}) as it should.

Finally, one observes that for a source dyon $(\bar{e},\bar{g})$ at the
origin of coordinates one has
\begin{eqnarray}
A^{\pm} &=& \frac{\bar{g}}{4\pi}(1 \pm \cos\theta) d\phi \ \ , \label{A} \\   
Z^{\pm} &=& \frac{\bar{e}}{4\pi}(1 \pm \cos\theta) d\phi  \label{Z} \ \ ,
\end{eqnarray}
which yields the quantization condition $e\bar{g} - g\bar{e}= 2\pi n \hbar$.

The preceding conclusions apply without change for all odd $p$-forms $A$. However, when $p$ is even,  the term (\ref{aa}) comes into the action with the opposite sign. This stems from the fact that
the string contribution to $^*\! F$ in (\ref{F}) is $^*(^*\! G)=(-1)^p G$,
and is what is responsible for the difference in sign in the expression for
$\pi_{\mu}$ for odd and even $p$ (compare (\ref{magnetic}) with
(\ref{qconstraint})). Thus for any $p$ the analog of (\ref{con})is given by
\begin{equation}
\omega_{\mu_{1}\cdots\mu_{p}} = e A_{\mu_{1}\cdots\mu_{p}} + (-1)^p g
Z_{\mu_{1}\cdots\mu_{p}} \ .
\label{gcon}
\end{equation}

Furthermore, since the form of the equations of motion for $F$ and $^*\! H$
is the same for all $p$,  
the dual connection $Z_\mu$ for the source dyon is always  obtained from the
corresponding $A_{\mu}$ by simply replacing  $\bar{e}$ by $\bar{g}$, just
as was the case in (\ref{A}) and (\ref{Z}).
Therefore (\ref{gcon}) gives rise to a quantization condition of symmetry
$(-1)^p$.

%%%%%%%%%%%%%%%%%%%%%%%%%%%%%%%%%%%%%%%%%%%%%%%%%%%%%%%%%%%%%%%%%%%%
%%%%%%%%%%%%%%%%%%%%%%%%%%%%%%%%%%%%%%%%%%%%%%%%%%%%%%%%%%%%%%%%%%%%

\end{document}